\documentstyle[11pt]{article}
\newcommand{\be}{\begin{eqnarray}}

\newcommand{\ee}{\end{eqnarray}}

\title{
	\begin{flushright}
	{\normalsize TPI--MINN--95--20 \\
	NUC--MINN--95/17--T \\
	HEP--MINN--95--1348 \\
        DOE/ER/40561--177--INT95--00--79\\
	August 1995 \\}
	\end{flushright}
\bf     Quantum Corrections to the Weizs\"acker--Williams
        Gluon Distribution Function at Small x
       }
\author{
	Alejandro Ayala, Jamal Jalilian-Marian and Larry McLerran \\
	{\small\it School of Physics and Astronomy,
	University of Minnesota, Minneapolis, MN 55455} \\
	Raju Venugopalan \\
	{\small\it Institute for Nuclear Theory,
	University of Washington,
	Seattle, WA 98195 } \\
       }

\date{}

\begin{document}
\setcounter{page}{0}
\maketitle
\thispagestyle{empty}
\begin{center}
{\bf Abstract}\\
\end{center}

\noindent
We compute the quantum corrections to the gluon distribution function in the
background of a non--Abelian Weizs\"acker--Williams field. These corrections
are valid to all orders in the effective coupling $\alpha_s \mu$, where $\mu^2$
denotes the average valence quark color charge squared per unit area. We find
$\ln(1/x)$ logarithmic corrections to the classical gluon distribution
function. The one loop corrections to the classical Weizs\"acker-Williams field
do not contribute to these singular terms in the distribution function. Their
effect is to cause the running of $\alpha_s$.
\vfill \eject

\section{Introduction}

In a recent work~\cite{jamal} we computed the Green's function in light cone
gauge $A^+=0$ for the small fluctuations about a background
Weizs\"acker--Williams gluon field. This background field is generated by the
valence quarks in a large $A$ nucleus. For small x partons, these valence
quarks constitute a static and well localized source of color
fields~\cite{larry1}. The average color charge squared per unit area of the
valence quarks is denoted by $\mu^2$ and it is of the order of $A^{1/3}$
fm$^{-2}$. The quantity
$\mu$ is the only dimensionful parameter in the theory and as a
result, the coupling constant will run as a function of it.

Previously, two of us~(McL.--V.) have argued that we can compute the gluon
distribution function from the light cone gauge Green's function~\cite{larry3}.
In the present paper we use that Green's function to compute the corrections
induced by quantum fluctuations on the Weizs\"acker--Williams distribution
function.

There are several reasons for this computation to be of interest. From the
practical point of view, we hope that an analysis which provides us with a
better understanding of the initial conditions for the evolution of partons
in the collision of heavy nuclei to form a quark--gluon plasma will help
establish a firm foundation~\cite{heribert} for partonic cascade models
simulating such collisions~\cite{klaus}. For an alternative approach to the
problem of initial conditions in heavy ion collisions, see
reference~\cite{sasha}.

{}From the theoretical point of view, we hope to understand the small x
behavior
of a nucleus starting from a QCD based approach. Let us recall that it is
believed that in the small x region, the gluon distribution function computed
perturbatively including leading logarithmic contributions in $\ln(1/x)$ for a
single nucleon behaves like~\cite{smallx}
\be
   \frac{dN}{dx} \sim \frac{1}{x^{1+C\alpha_{s}}} \nonumber.
\ee
The steep rise of the gluon distribution function for small x is sometimes
referred to as the Lipatov enhancement. It is obtained by solving the
Balitsky--Fadin--Kuraev--Lipatov (BFKL) kernel for the t--channel exchange of a
perturbative pomeron~\cite{delduca}. This behavior is also exhibited in a
hadron where the large x part of the hadron wavefunction is taken to be a heavy
quark--antiquark state. By applying Hamiltonian perturbation theory to this
state, it is possible to reproduce the kernel of the BFKL
equation~\cite{mueller} for the emission of a large number of soft gluons.

However, the rapid rise of the gluon distribution function with smaller values
of x is in conflict with unitarity when considering the hadron scattering total
cross section at asymptotically high energies~\cite{smallx}. Physically, this
violation of unitarity can be understood to result from ignoring effects that
arise due to the large density of partons at very small values of x~\cite{glr}.
When the density of partons is so large that neighboring partons overlap, the
t--channel picture of an independent parton cascade in x breaks down. The
former signals that at very small values of x, the picture in which the partons
do not interact with each other has to be modified in order to comply with the
Froissart bound~\cite{froissart} on the growth of cross sections at
asymptotically high energies. Although some work has been done in recent years
to include these ``higher twist'' effects in describing parton evolution at
high densities~\cite{glr,muellqiu,laenena}, more still remains to be done in
devising a quantitative mechanism to limit this growth.

The regime of high parton density is the screening regime. This screening is
presumably responsible for the shadowing phenomena observed in deep inelastic
scattering experiments off nuclei at small x. It can be addressed as a
collective or many body effect. It is precisely this many body problem of
parton interactions that we seek to address in our work. As outlined in
references~\cite{larry1}~\cite{larry3}~\cite{larry2}, our formalism provides
us, by means of a novel weak coupling approach, with a technique to solve the
many body problem of wee parton distributions in a large nucleus.

In this work, we will focus on one of the theoretical aspects of the problem
--the nature of the small x terms in the gluon distribution function of the
Weizs\"{a}cker--Williams fields generated by the valence quarks in the infinite
momentum frame. We do this by computing a formula for the gluon distribution
function which includes the effects of all orders in the parameter
$\alpha_s\mu$. Working in the weak coupling regime $\alpha_s\mu \ll k_t$, we
extract from this formula an expression for the distribution function in
perturbation theory to second order in $\alpha_s$. Although we approach the
problem in weak coupling, we will show that after the corrections are
considered, the series expansion parameter becomes $\alpha_s \ln(1/x)$ which
may be large for small x's. We will argue that this forces us to devise a
method to isolate and sum up the leading contributions to all orders in that
effective expansion parameter in order to compute the modification to the
zeroth order $1/x$ distribution function.

The outline of this work is as follows: In section 2, we briefly review the
basic aspects of the model in which we treat the nuclear valence quarks as
static sources of color charge as seen by small x partons. We also go briefly
through the formalism that allows us to compute the Green's function for the
small fluctuations equation and finish the section by writing the formula for
this Green's function. In section 3, we use this Green's function and exploit
its relation to the gluon density in order to compute the gluon distribution of
small x gluons. We also derive from this result a formula for the leading small
x terms of the distribution function in perturbation theory valid to second
order in $\alpha_s$. In section 4, we compute the corrections to the background
field induced by the fluctuation field and show that the only effect that this
introduces can be absorbed into the renormalization of the background field
(which is related to the renormalization of the coupling constant to one loop).
Details of the calculations in section 3 and section 4 are discussed
in appendix I and appendix II respectively. Finally we summarize our results in
section 5.

\section{The model}

In QCD, a hadron is a cloud of virtual particles with a rather complicated
structure. The picture gets simpler when we consider the hadron as a quantum
system composed of quasi real particles (partons) with lifetimes much larger
than the characteristic interaction times. This can be done in a reference
frame where the hadron has a large momentum~\cite{glr}. Partons with large
lifetimes can produce new partons carrying smaller fractions x of the initial
hadron's momentum. The small x partons will therefore densely populate the
hadron and see the rest of it with its longitudinal dimension Lorentz
contracted to a thin disk. In our model, we look at the small x partons in a
large $A$ nucleus ($x\ll A^{-1/3}$) where the high parton density allows us to
use weak coupling techniques. The rest of the nucleus consists of the valence
quarks which carry most of the nuclear momentum. They are described as a static
(recoilless) source of color charges, in a reference frame in which they move
with the speed of light (infinite momentum frame)~\cite{larry1}.

The problem is well suited to be described using light cone
variables~\cite{lightcone}
\be
    y^{0}, y^{3} \longrightarrow y^{\pm} = (y^{0} \pm y^{3})/\sqrt{2}
    \label{eq:lightconevar}
\ee

In order to compute ground state properties of the wee partons, we define a
partition function for the system. This partition function includes the sum
over a large number of color configurations. To simplify the problem, we
resolve the transverse space direction into cells which contain a large number
of valence quarks, or equivalently, a large number of color charges. This
allows us to treat the sum over color configurations classically~\cite{larry1}.
To write the average over the color charges, we introduce a Gaussian weight by
inserting into the path integral representation of the partition function the
term
\be
   \exp \left\{ - \frac{1}{2\mu^2} \int d^{2}x_{t}~ \rho^{2}(x) \right\} \, ,
   \label{eq:coloraverage}
\ee
where $\rho$ is the color charge density (per unit area) and the parameter
$\mu^2$ is the average color charge density squared (per unit area) in units of
the coupling constant $g$. The introduction of the partition function, where we
average over the sources of external charge, allows us to formulate the theory
as a many body problem with modified propagators and vertices.

We treat the system perturbatively and the first step is to solve the classical
equations of motion
\be
   D_{\mu}F^{\mu\nu} = g J^{\nu} , \; \;
   J^{\nu} = \delta^{+\nu} \rho(x^{+},x_{t}) \delta (x^-) \, ,
\ee
for which (working in the light cone gauge $A_{-}=-A^{+}=0$) there exists a
solution with $A_{+}=-A^{-}=0$. We require
\be
   A_{i}(x)=\theta(x^{-})\alpha_{i}(x_{t}) \, ,
   \label{eq:trans}
\ee
(hereafter, latin indices refer to transverse variables) and furthermore
$F^{ij}=0$. The latter condition implies that $\alpha(x_{t})$ is a pure gauge
transform of the vacuum~\cite{larry2},
\be
   \tau \cdot \alpha_{i} =\frac{i}{g} U(x_t)
    \nabla_{i} U^{\dagger}(x_t), \label{eq:alphau}
\ee
where $U(x_t)$ is a $SU(3)$ local gauge transformation whose
spatial dependence is only on the two--dimensional transverse space.
It is subject to the physical gauge condition
\be
   \nabla \cdot \left(U(x_t) \nabla U^{\dagger}(x_t)\right) =
   -ig^2 \rho \label{eq:ufield} \, .
\ee
The $x^+$ (light cone time) dependence of the charge density is a consequence
of the extended current conservation law
\be
   D_{\mu}J^{\mu}=0.
   \label{eq:gaugeinv}
\ee
The integration over the sources $\rho$ in equation~(\ref{eq:coloraverage})
may be written as
\be
   \int [dU] \exp \left( -\frac{1}{\mu^{2}g^{4}}{\bf Tr}
   \left( \nabla_i \cdot U \frac{1}{i} \nabla_i U^{\dagger}
   \right)^{2} \right),\label{eq:CA}
\ee
where we have ignored the Faddeev--Popov determinant. Note that the effective
coupling constant for this theory is $g^{2}\mu$ so that the expansion parameter
becomes $\alpha_s \mu / p_{t}$.

The Green's function can be computed from the relation
\be
   G^{\alpha\beta}_{ij}(x,x') = \int\frac{d\lambda}{\lambda - i \epsilon}
   \int\frac{d^{4}p}{(2\pi)^{4}}\delta(\lambda-p^{2})
   (A^{\alpha}_{i})_{\lambda}(x)(A^{\beta}_{j})^{\dagger}_{\lambda}(x')
   \label{eq:green}
\ee
and the gluon distribution function can be computed from the Green's function
by the relation
\be
   \frac{dN}{d^{3}k} =  i \frac{2k^{+}}{(2\pi)^{3}}
                       \sum_{\alpha , i} G^{\alpha \alpha}_{i i}
                       (x^{+}, \vec{k} ;x^{+}, \vec{k})\label{eq:distfunc} \, .
\ee
To relate the Green's functions to the distribution function by the above
relation, the former must be averaged over the external sources of color
charge.

Indeed, the distribution function, to all orders, is related by the above
expression to the fully connected two point Green's function. This Green's
function is given by the relation
\be
\langle \langle AA\rangle\rangle_{\rho}=\langle \langle A_{cl}\rangle
\langle A_{cl}\rangle + \langle A_{q}A_{q}\rangle \rangle_{\rho} \, .
\label{eq:schdy1}
\ee
In the above, $\langle A_{cl}\rangle$ is the expectation value of the classical
field to all orders in $\hbar$. It can be expanded as
\be
\langle A_{cl}\rangle = A_{cl}^{(0)}+A_{cl}^{(1)}+...... \, ,
\label{eq:schdy2}
\ee
where $A_{cl}^{(0)}$ is the solution discussed in Eqs.~(4)--(5).
The one loop correction to the classical field, $A_{cl}^{(1)}$, is computed in
section 4 of this paper. The term $\langle A_{q}A_{q}\rangle$ above is the
small fluctuations Green's function computed to each order in the classical
field. The symbol $<\cdots>_{\rho}$ indicates that we have to average over the
external sources of color charge with the Gaussian weight described
above.

{}From the above, it is clear that the zeroth order contribution is $\langle
A_{cl}^{(0)} A_{cl}^{(0)}\rangle_{\rho}$. This contribution is the QCD analog
of the well known Weizs\"acker--Williams distribution in classical
electrodynamics. The general form of the solution is given in
reference~\cite{larry2}. In the range of momenta $\alpha_{s}\mu\ll k_{t} \ll
\mu$, the zeroth order solution $A^{\mu}$ yields a distribution function that,
written in terms of $x\equiv k^{+}/P^{+}$, with $P^{+}$ the nuclear
longitudinal light cone momentum, looks like
\be
   \frac{1}{\pi R^{2}} \frac{dN}{dxd^{2}k_{t}} =
   \frac{\alpha_{s} \mu^{2} (N_{c}^{2} - 1)}{\pi^{2}}
   \frac{1}{xk^{2}_{t}} \label{eq:WWdist} \, .
\ee
The above is the well known Weizs\"acker--Williams distribution scaled by
$\mu^2\approx A^{1/3}$ fm$^{-2}$.

With this formalism at hand, we can proceed to compute the next order
contribution to the gluon distribution function. Our strategy is to compute the
small fluctuations correction to the classical equation of motion. Writing the
field in terms of its background and fluctuation parts
\be
    A^{\mu}(x)=B^{\mu}(x) + b^{\mu}(x),
\ee
we are able to express the equation obeyed by $b^{\mu}$ as~\cite{jamal}
\be
   \left( D(B)^2 g^{\mu \nu} - D^\mu (B) D^\nu (B) \right)
   b_{\nu} - 2 F^{\mu \nu} b_{\nu} = 0.
   \label{eq:SM}
\ee
where $B$ is the background field which according to equation~(\ref{eq:trans})
is non--vanishing only for its transverse components. $D^{\mu}(B)$ is the
covariant derivative with $B$ as the gauge field (notice that $D^{\pm} =
\partial^{\pm}$). As discussed in reference~\cite{jamal}, the set of
equations~(\ref{eq:SM}) can be unambiguously solved in the gauge $A^-=0$, and
by means of equation~(\ref{eq:green}) we can compute the Green's function for
the fluctuation fields in this gauge. To obtain the Green's function in the
gauge $A^+=0$ (light cone gauge), we perform a gauge transformation on the
Green's function in the $A^-=0$ gauge and obtain finally (in the matrix
representation)
\be
G_{ij}^{\alpha\beta ;\alpha^\prime \beta^\prime}(x,y)
\!\!\!\!\!& = &\!\!\!\! - \! \int \!\!{{d^4p}\over {(2\pi)^4}}
{{e^{ip(x-y)}}\over {p^2-i\epsilon}}
\Bigg\{\! \bigg[\delta_{ij}\!+\!{p_ip_j\over {p^-p^+}}(2e^{ip^+(x^-\! - y^-)}
\!-\! e^{-ip^+y^-}\!-\! e^{ip^+x^-})\bigg]\nonumber \\
&\times &   \bigg[\theta(-x^{-})\theta(-y^{-})\tau_{a}^{\alpha\beta}
\tau_{a}^{\alpha^\prime \beta^\prime} + \theta(x^-)\theta(y^-)\,
F_{a}^{\dagger\alpha\beta}(x_{t})
F_{a}^{\dagger\alpha^\prime \beta^\prime}(y_{t})
\bigg]\nonumber \\
\!\!\!\!\!&+ &\!\!\!\!
 \theta(-x^-) \theta(y^-)
\int \!{{d^2 q_t} \over {(2\pi)^2}} d^2 z_t \,\, e^{i(q^{+}-p^{+})y^{-}}
e^{i(p_t-q_t)(y_t-z_t)} \nonumber \\
\!\!\!\!\!&\times &\!\!\!\!
F_{a}^{\dagger\alpha \beta}(z_{t})
F_{a}^{\dagger\alpha^\prime  \beta^\prime}\!(y_{t})\bigg[\delta_{ij} +
\frac{p_{i}p_{j}}{p^-p^{+}}
(e^{ip^{+}x^{-}}\!-1)\nonumber \\
\!\!\!\!\!& + &\!\!\!\!\!\!   \frac{q_{i}q_{j}}{p^-q^{+}}
(e^{-iq^{+}y^{-}}\!-1)
+ \frac{p_{i}q_{j}p_{t}\cdot q_{t}}
{(p^{-}p^{+})(p^{-}q^{+})}(e^{ip^{+}x^{-}}\!-1)
(e^{-iq^{+}y^{-}}\!-1)\bigg]\nonumber \\
\!\!\! & + &\!\!\!\!\!
\theta(x^-) \theta(-y^-)
\int\! {{d^2 q_t} \over {(2\pi)^2}} d^2 z_t \,\,
 e^{i(p_t-q_t)(z_t - x_t)} e^{-i(q^{+}-p^{+})x^{-}}\nonumber \\
\!\!\!\!\!&\times &\!\!\!\! F_{a}^{\dagger\alpha \beta}(x_{t})
F_{a}^{\dagger\alpha^\prime  \beta^\prime}\!(z_{t})\bigg[\delta_{ij} +
\frac{p_{i}p_{j}}{p^{-}p^{+}}(e^{-ip^{+}y^{-}}\!\!-1)
 +  \frac{q_{i}q_{j}}{p^{-}q^{+}}(e^{iq^{+}x^{-}}\!-1)\nonumber \\
\!\!\!\!\!& + &\!\!\!\!\!\!
\frac{q_{i}p_{j}p_{t}\cdot q_{t}}{(p^{-}p^{+})(p^{-}q^{+})}
(e^{-ip^{+}y^{-}}\!-1)
(e^{iq^{+}x^{-}}\!-1)\bigg]\Bigg\} \, ,
\label{eq:GF}
\ee
where $q^{+}=p^{+}+ \frac{q_{t}^{2}-p_{t}^{2}}{2p^{-}}$ and
\be
   F_{a}^{\dagger\alpha \beta}(x_{t}) =\bigg( U(x_{t})\tau_{a}U^{\dagger}
   (x_{t})\bigg)^{\alpha\beta}.
\ee
The above result was derived in reference~\cite{jamal}. The regularization of
the poles in $p^-$ and $p^+$ is a rather subtle issue and was discussed at some
length in the above mentioned paper.

The Green's function $G_{ij}(x,y)$ contains the physical degrees of freedom
that we need to relate to the gluon density. In the following section, the
above small fluctuations propagator will be used to compute corrections to the
Weizs\"acker--Williams distribution function. We will show that logarithmic
corrections, both in x and $k_t$, arise from here. In section~4, we discuss the
corrections that the small fluctuations induce on the classical field to one
loop. These corrections, in contrast, provide no such logarithmic terms to the
distribution function and their only effect is to renormalize the coupling
constant and the background field.

In appendix I, we compute the Fourier transform of the Green's function.
This result is useful for the computation of the distribution functions
performed in the next two sections.

\section{The gluon distribution function}

In this section, we will show how one can obtain the gluon distribution
function from the Green's function~(equation~\ref{eq:GF}), corresponding to the
small fluctuations about the Weizs\"{a}cker--Williams background color field.
We will be concerned with the structure of the leading terms for small x and at
the end of the section compute a formula for the gluon distribution function
valid to $\alpha_{s}^{2}$. To compute the distribution function we need to sum
over all possible color configurations. This color average involves the two
dimensional gauge fields $U$. We start by recalling the properties of these
gauge fields under our color averaging.

\subsection{Correlation functions involving the gauge transformations $U$}

According to equation~(\ref{eq:alphau}) the gauge transformations $U(x_t)$
carry the information on the background field which enters the Green's
function~(\ref{eq:GF}). These gauge transformations have the interesting
property that the color average with the Gaussian weight (defined by
equation~(\ref{eq:CA})), of the combination
\be
   U^{\dagger}(x_t)\tau^{a}U(x_t)U^{\dagger}(y_t)\tau^{a}
   U(y_t)
\ee
can be written as~\cite{larry3}
\be
<\! Tr U^{\dagger}(x_t)\tau^{a}U(x_t)U^{\dagger}(y_t)\tau^{a}U(y_t)\!>=
\!\frac{(N_c^2 -1)}{2}\!\Gamma (x_{t}\!-\!y_{t}) \label{eq:correl}
\ee
where $N_c$ is the number of colors and summation over repeated indices is
implied.

As pointed out in reference~\cite{larry1}, the average over the color sources
yields the information about the ground state properties of the system. The
average with the Gaussian weight is an artifact that simplifies the computation
and we expect that as long as we resolve the nucleus on a transverse size
much larger than the typical transverse quark separation, such an artifact
is justified.

The function $\Gamma$ factorizes the dependence on the transverse
coordinates $x_t$,~$y_t$ and is a function of their
difference. Moreover, from equation~(\ref{eq:correl}) we see that
$\Gamma$ is real and also
\be
   \Gamma(0)=1.
   \label{eq:gammaofzero}
\ee
Defining the Fourier transform of $\Gamma(x_t)$
\be
   \gamma(p_t) = \int d^{2}x_t e^{-ip_t x_t}\Gamma(x_t)
   \label{eq:Foutransgamma}
\ee
we have, together with~(\ref{eq:gammaofzero}), the sum rule
\be
   \int \frac{d^2 p_t}{(2\pi)^2} \gamma(p_t) = 1.
   \label{eq:sumrule}
\ee

The color charge at a given transverse location will be zero on average and the
only way to generate a non--zero color charge will be by fluctuations.
Equation~(\ref{eq:CA}) can be thought of as the generator of those
fluctuations and thus the function $\Gamma(x_t, y_t)$ represents the
correlator of fluctuating fields at the transverse locations $x_t$ and
$y_t$.

In momentum space, the function $\gamma(p_t)$ can be formally computed by
expanding the exponential in~(\ref{eq:CA}) in powers of the coupling parameter
$\alpha_{s}\mu / p_t$ (weak coupling regime). This was done in
reference~\cite{larry3} for scalars. For gluons, the result for $\alpha_s\mu
\ll p_t$ is
\be
   \gamma(p_t) = (4\pi)^2\frac{\alpha_{s}^{2}\mu^2}{p_{t}^{4}}N_c\,.
   \label{eq:gammaweak}
\ee
We notice that the expansion is only necessary in order to analytically compute
expressions of the form
\be
   \int d^{2}p_t f(p_t)\gamma(p_t) \, ,
   \label{eq:example}
\ee
with $f(p_t)$ a non--trivial function of $p_t$. However, in principle, we can
perform a numerical analysis to take into account the many possible different
configurations of the external field contributing to expressions such
as~(\ref{eq:example}). This is equivalent to considering the effect to all
orders in $\alpha_{s}\mu / p_t$ of the different configurations of the
background field. For a quantitative discussion about the properties of the
distribution function, we will restrict ourselves to the weak coupling regime
for which $\gamma(p_t)$ is given by equation~(\ref{eq:gammaweak}).

\subsection{The distribution function}

With the above remarks in mind, we proceed to the computation of the gluon
distribution function. We use the formula for the distribution function
\be
   \frac{dN}{d^{3}k} =  i \frac{2k^{+}}{(2\pi)^{3}}
   \lim_{k^+\rightarrow {k'}^+}\int \frac{dk^-}{2\pi}
   \int \frac{d{k'}^-}{2\pi} <D^{aa}_{ii}(k,k')\!\!> \, ,
\ee
where $<D^{aa}_{ii}(k,k')\!\!>$ is the small fluctuations propagator in
momentum space, traced over the color and Lorentz indices and averaged over the
external sources of color charge. This formula follows from computing
$<a^\dagger(p) a(p)>$ as an expectation value for the gluon Fock space
distribution function in the ground state generated by the external valence
charges~\cite{larry3}.

In appendix I, we derive explicitly an expression for
$<D^{aa}_{ii}(k,k')\!\!>$. Using this result, we obtain the following integral
expression for the distribution function
\be
{1 \over {\pi R^2}} {{dN} \over {d^3k}}
& = &
{{2ik^+} \over {(2\pi)^3}}(N_c^2-1)
\lim_{k^+ \rightarrow k^{\prime +}}
\int
{{dp^+d^2p_tdk^-} \over {(2\pi)^4}}
\nonumber \\
& &
\Bigg\{
{1 \over {p_t^2-2p^+k^--i\epsilon}}
\Bigg(
2 +
{p_t^2 \over {k^-k^+k^{\prime +}}}
(2p^+-k^+-k^{\prime +})
\Bigg)
\nonumber \\
&\times&
\Bigg(
-(2\pi)^2 \delta^{(2)} (p_t-k_t)
{1 \over {p^+-k^++i\epsilon}}
{1 \over {p^+-k^{\prime +} - i\epsilon}}
\nonumber \\
&-&
\gamma(p_t-k_t)
{1 \over {p^+-k^+-i\epsilon}}
{1 \over {p^+-k^{\prime +} + i\epsilon}}
\Bigg)
\nonumber \\
&+&
\gamma(p_t-k_t)
{1 \over {k_t^2-2p^+k^- - i\epsilon}}
\Bigg(
 2 -
{{k_t^2+p_t^2} \over {k^-k^+}} +
{{(p_t \cdot k_t)^2} \over {(k^{-2}k^{+2})}}
\Bigg)
\nonumber \\
&\times&
\Bigg(
{1 \over {p^+-k^+ - i\epsilon}}
{1 \over {q^+-k^+ - i\epsilon}}+
{1 \over {p^+-k^+ + i\epsilon}}
{1 \over {q^+-k^+ + i\epsilon}}
\Bigg)
\Bigg\}
\ee
In the last term of this equation, we have taken the limit that $k^+
\rightarrow k^{\prime +}$ since this term has no singularity in that limit.

We now do the integral over $k^-$.  We assume that
$k^+ > 0$. When we do the integral, two classes
of terms result.  The first set of terms arise from the explicit
$k^-$ dependence in the above equation and are non-zero.  The second
set of terms arise from the $q^+$ in the last terms of the above equation.
These terms result in an unrestricted integral over $p^+$.
One can show that all the singularities of the resulting integrand
are on the same side of the $p^+$ integration contour in the complex
$p^+$ plane.  They therefore integrate to zero.  (There is a possible
ambiguity in the closing of contours associated with the contour at
infinity, but this term does not have any contribution proportional to
$\ln(1/x)$.)  Therefore, we only get the contribution from the first term,
which is only nonzero for $p^+ < 0$,
\be
{1 \over {\pi R^2}} {{dN} \over {d^3k}} & = &
(N_c^2-1){{4k^+} \over {(2\pi)^3}}
\int^0_{-\infty} {{dp^+d^2p_t} \over {(2\pi)^3}}
\nonumber \\
&\times&
\Bigg\{ {1 \over {p^+(p^+-k^+)(k^++p^+p_t^2/k_t^2)}}
\left[\gamma(p_t-k_t) - (2\pi)^2\delta^{(2)} (p_t-k_t)\right]
\nonumber \\
&\times&
\Bigg[ 1 - {p^+ \over k^+} \left( 1 + {p_t^2 \over k_t^2} \right)
+2 \left( {p^+ \over k^+} \right)^2 { (p_t \cdot k_t)^2 \over k_t^4}
\Bigg] \Bigg\}\, .
\ee

Now in this expression, we shall only be concerned with those terms which are
proportional to $\ln(1/x)$.  The terms not proportional to $\ln(1/x)$ are
non-leading for small x.  Moreover, we have found that within our approach,
these terms are inherently ambiguous. This is due to the fact that the
$\ln(1/x)$ terms can only arise by regulating the singularity in the above
integral as $p^+ \rightarrow \infty$. We do this by making the upper limit of
integration, to be  of the same
order than the total momentum of a typical nucleon in the nucleus.  Of course
different regularization schemes will affect the non--leading terms in
different
ways. Presumably, the detailed longitudinal structure of the valence quark
charge distribution must be known before these terms may be evaluated.

After some straightforward algebra, we find
\be
\Bigg(\frac{1}{\pi R^{2}}\frac{dN}{dxd^{2}k_{t}}\Bigg)_q\!\!\!
&=&\!\!\! \frac{8(N^{2}_{c}-1)}{(2\pi )^{4}}
\frac{1}{x}\int {{d^2 p_t }\over {(2\pi )^2}}
\gamma (p_t -k_t ) \bigg[ 1 -
\frac{(p_t \cdot k_t )^2}{p_t^2 k_t^2}\bigg]\ln \left(\frac{1}{x}\right)
\label{eq:MAINRES}
\ee
where the subindex $q$ in the left hand side of the above equation refers to
the correction to the distribution function from the small quantum fluctuation
field. Equation~(\ref{eq:MAINRES}) is our main result. It is normalized so that
the vacuum density is zero. This can be checked by setting $U=1$ in the above
equation.

The terms above can be written in the form
\be
   \Bigg(\frac{1}{\pi R^{2}}\frac{dN}{dxd^{2}k_{t}}\Bigg)_{q}\!\!=
   \left(\frac{N_c^2-1}{2\pi^4}\right)\frac{1}{x}\int \frac{d^2p_t}{(2\pi)^2}
   \gamma(p_t-k_t)
   \Bigg\{ \frac{p_{t}^{2}k_{t}^{2} - (p_t \cdot k_t)^2}
                {p_{t}^{2}k_{t}^{2}}
   \Bigg\} \ln \left(\frac{1}{x}\right) .
   \label{eq:secandthird}
\ee
Now, shift the variable of integration $p_t \rightarrow p_t+k_t$ and expand
$\gamma(p_t)$ in weak coupling. This restricts the lower limit of integration
for the radial component of $\vec{p_t}$ to be $\alpha_s\mu$, which comes from
the weak coupling expansion of $\gamma$. Thus the above expression becomes
\be
   \Bigg(\frac{1}{\pi R^{2}}\frac{dN}{dxd^{2}k_{t}}\Bigg)_{q}&=&
   8N_c(N_c^2-1)\frac{\alpha_{s}^{2}\mu^2}{(\pi)^2}\frac{1}{x}\nonumber \\
   &\times&\int_{0}^{2\pi} d\theta \int_{\alpha_s\mu}^{\infty}
   dp_t \frac{(1-\cos^2\theta)}
             {p_t[p_{t}^{2}+k_{t}^{2}+2p_tk_t\cos\theta]}
   \ln \left(\frac{1}{x}\right)
   \label{eq:interval}
\ee
where $\theta$ is the angle between $\vec{p_t}$ and $\vec{k_t}$.
The integral above can be performed exactly and the contribution
to~(\ref{eq:distfunc}) from the $\ln (1/x)$ terms becomes
\be
   \Bigg(\frac{1}{\pi R^{2}}\frac{dN}{dxd^{2}k_{t}}\Bigg)_{q1}=
   N_c(N_{c}^{2}-1)
\frac{\alpha_{s}^{2}\mu^{2}}{xk_{t}^{2}}\frac{C(k_t)}{\pi^3}
   \ln \left(\frac{1}{x}\right).
   \label{eq:contsecandthird}
\ee
with $C(k_t)$ given by
\be
   C(k_t)=
   2\left( \ln (\frac{k_t}{\alpha_s\mu}) + \frac{1}{2}\right).
   \label{eq:coeffitientCk}
\ee

Notice that the above expression means that the second term in the
perturbative expansion of $dN/dxd^2k_t$ in $\alpha_s$,
develops the large factor $\ln(1/x)$ and that in the kinematical region of
interest, the product $\alpha_s\ln(1/x)$ is not small.

Furthermore, let us impose ordering in transverse momentum. This is the
statement that the main contribution to the distribution function comes from
the momentum region for which the emitted gluon's transverse momentum is larger
than that of the original one~\cite{glr}. The
effect is to restrict the integration interval for the radial component of
$\vec{p_t}$ in equation~(\ref{eq:interval}) which now runs between
$\alpha_s\mu$ and $k_t$. The reader can check that the above results in the
modification of~(\ref{eq:coeffitientCk}) which now reads like
\be
   C(k_t)=
   2\ln \left(\frac{k_t}{\alpha_s\mu}\right).
   \label{eq:coeffitientC}
\ee

We can now include the contribution from the background
Weizs\"{a}cker--Williams  field as given by equation~(\ref{eq:WWdist}). Thus
finally, the perturbative expression for the gluon distribution function to
second order in $\alpha_s$ becomes
\be
   \frac{1}{\pi R^{2}} \frac{dN}{dxd^{2}k_{t}}& = &
   \frac{\alpha_{s} \mu^{2} (N_{c}^{2} - 1)}{\pi^{2}}
   \frac{1}{xk^{2}_{t}} \Bigg\{ 1 + \frac{\alpha _s N_c}{\pi} C(k_t)
   \ln \left(\frac{1}{x}\right)  \Bigg\}.
   \label{eq:gluondistfuncsec}
\ee

Equation~(\ref{eq:gluondistfuncsec}) contains both $\ln(1/x)$ and $\ln(k_t)$
corrections to the $1/(xk_t^2)$ distribution and they represent the first order
contributions to the perturbative expansion for the distribution function. In
the kinematical region of validity, these corrections are large. This signals
that in order to properly account for the perturbative corrections one has to
devise a mechanism to isolate and sum up these leading contributions. Also
notice that equation~(\ref{eq:MAINRES}) is more general. In principle, it
contains the information about the non--perturbative corrections as well. That
information is in the function $\gamma(p_t)$ and it can be extracted by means
of a Monte Carlo analysis for the whole $k_t$ domain. These issues will be
treated in a future work.

Diagrammatically, we can represent the background gluon field coupled
to the external source (valence quarks), in momentum space, by means of
figure 1a. The background field (wavy line) is by itself of order $1/g$,
according to equation~(\ref{eq:alphau}) and the coupling to the external source
(cross) can be considered to n-th order in the parameter $g^2\mu/k_t$ by means
of the weak coupling expansion of equation~(\ref{eq:CA}). As an example, the
Weizs\"{a}cker--Williams distribution is obtained through the correlation of
the
background field taking the average over the source to first order (n=1) in
$g^2\mu/k_t$. This can be represented as in figure 1b where the broken wavy
line means that the momentum $k$ is not integrated over.

The gluon propagator in the presence of the background field can be computed
perturbatively in the coupling constant $g$ and the m-th order gluon propagator
can be represented as in figure~2a. This is because the perturbative expansion
of the gluon field involves its coupling to the background field through the
covariant derivative and the background field acts as the gauge field. Notice
that m has to be even since the gluon propagator is the correlator of two gluon
fields and each time we couple the background field to the gluon field we
introduce one power of $g$. In particular, the gluon propagator to second order
in $g$ can be represented as in figure 2b. As suggested in this figure, the
explicit dependence on $\alpha_s$ of a quantity such as the gluon distribution
function (which involves the gluon propagator) comes about only after
performing our color average through the expansion in $g^2\mu/k_t$. This is
because the coupling constant dependence of the background field and the order
of the perturbative expansion offset each other. As shown in section 3,
when computing the leading small x terms for the gluon distribution function,
any term for which we can use the sum rule~(\ref{eq:sumrule}) will not exhibit
an explicit coupling constant dependence. This becomes the criterion to decide
that such terms are vacuum contributions.

The gluon distribution function computed in section 3 can be represented by the
diagram in figure 3, where we expanded the coupling with the external source
to first order in $g^2\mu/k_t$.

\section{Loop corrections to the classical field}

Thus far, we have been concerned exclusively with the contribution of the small
fluctuations propagator to the gluon distribution function. We have shown that
this propagator induces large corrections proportional to $\alpha _{s}\ln(1/x)
\ln(k_t^2)$
and $\alpha _{s}\ln(1/x)$ to the distribution function and have argued
that the  presence of these large logarithms signals the need to devise a
method to sum them up to all orders in the perturbative regime. Before we do
that we need to consider another contribution, to the same order, which comes
from the corrections to the lowest order classical field induced by quantum
fluctuations (see figure 4). This is apparant from equation~(\ref{eq:schdy1})
and equation~(\ref{eq:schdy2}) where one sees that there is a contribution
$\langle
A_{cl}^{(1)} A_{cl}^{(0)}\rangle_{\rho}$ of the same order as $\langle \langle
A_{q} A_{q} \rangle \rangle_{\rho}$.

In this section, we will compute the correction to the lowest order classical
field induced by the quantum fluctuations. We will start by writing the total
field $A^{\mu}$ in terms of background (classical) and fluctuation
(quantum) pieces allowing for the possibility that the background field may now
be different from our lowest order classical (Weizs\"acker--Williams) solution.
We will then write the equations of motion in terms of these new background
and fluctuation fields keeping terms up to and including second order in the
fluctuation fields.

Our strategy will be to consider the expectation value of the equations of
motion (in the path integral sense) and to relate the correlator of two quantum
fields to the gluon propagator in equation~(\ref{eq:GF}). We will show that
only the $+$ component of the equations of motion is modified and that the
change could be thought of as the appearance of an induced current generated by
the loop of fluctuation fields. We then proceed to explicitly compute this
induced current and show that its effect is to renormalize the coupling
constant $g$ and the original background field. In other words, the
modification induced by quantum fluctuations on the classical equations of
motion can be cast into the standard expression for the renormalization of the
coupling constant and the original background field to one loop in the light
cone gauge. This result in itself is not surprising to QCD practitioners (
see for instance reference~\cite{doksh}). What is surprising is that
this result
persists to all orders in the effective coupling $\alpha_S\mu$.

We start with the classical equations of motion
\be
   D_{\mu}F^{\mu\nu}_a = g J^{\nu}_a
   \label{eq:classical}
\ee
and expand the full gluon field as
\be
   A^{\mu} = B^{\mu} + b^{\mu}
\ee
where $B^{\mu}$ is the background (classical) field, that is
$<A^{\mu}>\, =B^{\mu}$ while $b^{\mu}$ is the fluctuation (quantum) field
with $<b^{\mu}>\, =0$. Keeping up to quadratic terms in $b^{\mu}$,
the $+$ component of the equations of motion can be written as
\be
   \partial_{-}\partial_{-}B_{a}^{-}
   +(D_i\partial_- B^i)_{a} = gj_a^+ + g<J^{+}_{a}>
   \label{eq:plus}
\ee
where $j_a^+(x)= f_{abc}<b_{b}^{i}(x)\partial^+b_{c}^{i}(x)>$. Also, $D^{\mu}$
is the covariant derivative with $B^{\mu}$ as the gauge potential.
The corresponding expressions for the minus and transverse components
of the equations of motion look like
\be
   (D_iD^iB^-)_{a} - (D_i\partial^-B^i)_{a} +
   (D_+\partial^+ B^-)_a + \nonumber \\
   gf_{abc}\Bigg\{ <b_+^b\partial^+b_c^->+(D_i<b^ib^->)_{bc}+\nonumber \\
                   <b_b^i(D^ib^-)_c>-<b_b^i(D^-b^i)_c>
           \Bigg\} = 0
   \label{eq:minus}
\ee
\be
   (D_jD^jB^i)_a - (D_j\partial^iB^j)_a+(D_+\partial^+ B^i)_a -
   (\partial_-D^iB^-)_a + \nonumber \\
   (\partial_{-}\partial^- B^i)_a + gf_{abc} \Bigg\{
   <b_+^b\partial^+b_i^c>+\partial_-<b_b^-b_c^i> +
   \nonumber \\
   (D_j<b^jb^i>)_{bc} - <b_b^j(D^ib^j)_c> + <b_b^j(D^jb^i)_c> \Bigg\} = 0.
   \label{eq:transverse}
\ee
The expectation values of bilinear products
of fields are related to the gluon propagator by the relation
\be
   <b_a^{\mu}(x)b_b^{\nu}(y)> = -iG_{ab}^{\mu\nu}(x,y)\, .
   \label{eq:relgreenexp}
\ee
In the above, $a,b,c$\ldots are color indices and $\mu$, $\nu$ are Lorentz
indices with $i,j$ representing the transverse components.

The reader may verify that all the terms involving bilinear products of $b^\mu$
in the minus and transverse components of the equations of motion either vanish
by explicit computation, or, because they are symmetric in the color indices
$b$ and $c$, obviously do not contribute since they are always contracted with
the totally antisymmetric structure constants $f_{abc}$. In other words, the
minus and transverse components of the classical equations of motion are not
modified by the quantum fluctuations and the set of equations reduces to
\be
   -\partial_{-}\partial^{+}B_{a}^{-}
   -(D_i\partial^+B^i)_{a}& = &
   gj^+_a + g<J^{+}_{a}>,\nonumber \\
   (D_iD^iB^-)_{a} - (D_i\partial^-B^i)_{a} +
   (D_{+}\partial^+B^{-})a & = & 0, \nonumber \\
   (D_jD^jB^i)_a - (D_j\partial^iB^j)_a + (D_+\partial^+B^i)_a -
   (\partial_-D^iB^-)_a & = & 0.
   \label{eq:set}
\ee
{}From now on we will concentrate only on the plus component of the equation
of motion given by equation~(\ref{eq:plus}). It is clear that
this equation is modified by the quantum fluctuations due to the presence
of the induced current. In order to understand this effect, we need to
evaluate this term explicitly. For this purpose, we write it in the following
way
\be
 j^+_a(x)=f_{abc}<b_{b}^{i}(x)\partial^+b_{c}^{i}(x)>=
 i f_{abc}\lim_{y\rightarrow x} \frac{\partial}{\partial y^-}
   G_{bc}^{ii}(x,y).
 \label{eq:term}
\ee
Diagramatically, this term can be represented as in figure~4 where the wavy
line is the background field and the spiral represents the loop of the
fluctuation field. The loop is the vacuum polarization tensor and the component
which contributes to the induced current and modifies the background field
(which is purely transverse) is $\Pi^{+i}$. This  allows us to represent the
term $\partial^+G^{ii} \equiv D^+G^{ii}$ as $\Pi^{+i}B^i$.

We now proceed to compute the induced current explicitly. We will use
our expression for the gluon propagator as given by equation~(\ref{eq:GF}).
The first observation is that the terms in the Green's function with both
$x^-$ and $y^-$ negative will be symmetric in the color indices
and will not contribute. Also, it can be shown that the terms with both $x^-$
and $y^-$ positive yield (after we implement the limit $y \rightarrow x$ in a
Lorentz covariant way) an infinite constant (independent of the transverse
loop momentum) which vanishes  upon dimensional regularization~\cite{kaku}.

However, the terms in the Green's function with opposite signs of
$x^-$ and $y^-$ are a bit tricky. For these terms taking the partial
derivative with respect to $y^-$ followed by the limit $y \rightarrow x $
is a very delicate operation and must be performed carefully. We find it more
convenient to rewrite the terms with opposite signs of $x^-$ and $y^-$ in the
Green's function in  such a way as to avoid acting with $\partial/\partial y^-$
on the terms $\theta(\pm y^-)$. To do so, we will change the two dimensional
integral over $q_t$ to a four dimensional integral over $q$. We can show
that as a result, the product of theta functions of $x^-$ and $y^-$ will be
replaced by theta functions of the light cone energy $p^-$. After  some
long but straightforward algebra we can rewrite the terms with opposite signs
of $x^-$ and $y^-$ in the Green's function (which we call $D^{bc}_{ij}$) as
\be
   D_{ij}^{bc}(x,y)\equiv
   (2i)\int\frac{d^4p}{(2\pi)^4}\frac{d^4q}{(2\pi)^4}(2\pi)
   \delta(p^- - q^-)(p^- + q^-)
   \frac{e^{ip(x-y)}}{(p^2-i\epsilon)(q^2-i\epsilon)}\nonumber \\
   \times\Bigg\{\theta(p^-)e^{-i(p-q)x}Tr[U^{\dagger}(x_t)\tau^bU(x_t)
                                         \tilde{F}^c(p_t-q_t)]\times
   \Bigg[\delta_{ij} +\frac{p_ip_j}{p^-p^+}(e^{-ip^+y^-}-1)\nonumber \\
		     +\frac{q_iq_j}{q^-q^+}(e^{iq^+x^-}-1)
         +\frac{q_ip_j(q_t\cdot p_t)}{p^-p^+q^-q^+}
          (e^{-ip^+y^-}-1)(e^{iq^+x^-}-1)\Bigg]\nonumber \\
          -\theta(-p^-)e^{i(p-q)y}Tr[\tilde{F}^b(q_t-p_t)
                                       U^{\dagger}(y_t)\tau^cU(y_t)]\times
   \Bigg[\delta_{ij}  +\frac{q_iq_j}{q^-q^+}(e^{-iq^+y^-}-1)\nonumber \\
                      +\frac{p_ip_j}{p^-p^+}(e^{ip^+x^-}-1)
          +\frac{p_iq_j(q_t\cdot p_t)}{p^-p^+q^-q^+}
          (e^{-iq^+y^-}-1)(e^{ip^+x^-}-1)\Bigg]\Bigg\}\label{eq:crossterms}
\ee
where $\tilde{F}^b(p_t)$ is the Fourier transform of
$F^b(x_t)\equiv U^{\dagger}(x_t)\tau^bU(x_t)$. In the above expression, we are
working with the color components of expression~(\ref{eq:GF}) (as opposed to
the matrix notation), which is more suitable for the computation at hand.

According to equation~(\ref{eq:term}), we now have to compute
\be
   j^+_a(x)=if_{abc}\lim_{y\rightarrow x}\frac{\partial}{\partial y^-}
   D_{ii}^{bc}(x,y).
   \label{eq:renorm}
\ee
There are three distinct pieces to the computation corresponding to the
different number of factors of $p^-$ ($q^-$) in the denominator of each
term in expression~(\ref{eq:crossterms}). For the rest of the section, we will
outline the procedure for the computation of one of them, namely, the term
proportional to $\delta_{ij}$ and quote the result for the other two terms.
In appendix II, we will show the detailed computation of some of the
integrals necessary to fill in the intermediate steps.

There are two important details to keep in mind while computing explicitly
expression~(\ref{eq:renorm}): first, we have to implement the limiting
procedure in a Lorentz covariant way. Second, when carrying out the integration
over the relative and total (light cone) energies, we have to allow for a
non--zero energy flow into the loop by not setting the total energy to zero, in
spite of the structure of the integral which seems to require it to be so. The
procedure is nothing but the well known {\it point splitting method}.

In order to implement the limit $y\rightarrow x$ in a Lorentz covariant
way, we first transform the induced current to momentum space and then
integrate over the relative (loop) momentum. We make use of the following
identity. Let $f(x,y)$ be a function of the four--dimensional variables $x$ and
$y$ for which we want to compute the limit when $x \rightarrow y$.
Fourier transform $f$ to momentum space with respect to both $x$ and $y$
\be
\tilde{f}(k,k')=   \int d^4xd^4y~f(x,y)
   e^{ikx}e^{ik'y}
\ee
and then make the change of variables
\be
   s=\frac{k-k'}{2}\nonumber \\
   S=\frac{k+k'}{2}\label{eq:transformation}
\ee
where $s$ and $S$ can be thought of as the relative and total momenta
respectively. Then
\be
   \tilde{f}(s,S)=   \int d^4xd^4yf(x,y)
   e^{is(x-y)}e^{iS(x+y)}.
\ee
Integrating over $d^4s/(2\pi)^4$ will give $\delta^4(x-y)$
which will set $y=x$ upon integrating over $y$. Hence,
\be
   \int d^4s \tilde{f}(s,S)=   \int d^4x~e^{i2Sx}f(x,x)\, ,
\ee
which is the expression for the Fourier transform of $f(x,y)$ in the limit when
$y \rightarrow x$ as a function of $2S$. With the above remarks in mind, we
proceed to take the partial derivative with respect to $y^-$ in
equation~(\ref{eq:crossterms}) and then to Fourier transform with respect to
$x$ and $y$ to the momentum variables $k$ and $k'$. Let us look at the piece
proportional to $\delta_{ij}$. Set $i=j$ and then perform the $p^{\pm}$,
$q^{\pm}$ and $p_t$ integrations to get
\be
   \int d^4xd^4ye^{ikx}e^{ik'y}\frac{\partial}{\partial y^-}
   D^{ii}_{bc}(x,y)_{(1)}&\!\!\!=\!\!\!&
   (-2\pi)\delta({k'}^- + k^-)
   \int \frac{d^2q_t}{(2\pi)^2}\nonumber \\
   &\!\!\!\times\!\!\!&
   \frac{({k'}^- - k^-){k'}^+}{({k'}^-)(k^-)}
   Tr[\tilde{F}^a(k_t+q_t)\tilde{F}^b({k'}_t-q_t)]\nonumber \\
   &\!\!\!\times\!\!\!&
   \left\{
   \frac{\theta({k'}^-)}{({k'}^+ - \frac{({k'}_{t}^{2} -i\epsilon )}{2{k'}^-})
   (k^+ - \frac{(q_{t}^{2} -i\epsilon )}{2k^-})}\right.\nonumber \\
   &\!\!\!-\!\!\!& \left.
   \frac{\theta({-k'}^-)}{({k}^+ - \frac{({k}_{t}^{2} -i\epsilon )}{2{k}^-})
   ({k'}^+ - \frac{(q_{t}^{2} -i\epsilon )}{2{k'}^-})}\right\},
\ee
where the index $1$ refers to our considering the first of the terms in
equation~(\ref{eq:crossterms}), namely, the term proportional to $\delta_{ij}$.
Now, perform the change of variables~(\ref{eq:transformation}) to write the
above expression in terms of relative and total momentum and integrate over
the relative momentum. The expression to evaluate becomes
\be
   \int \frac{d^4s}{(2\pi)^4}
   d^4xd^4ye^{is(x-y)}e^{iS(x+y)}
   \frac{\partial}{\partial y^-}
   D^{ii}_{bc}(x,y)_{(1)}=
   (-2\pi)\delta(2S^-)\nonumber \\
   \times \int \frac{d^4s}{(2\pi)^4}
   \frac{d^2q_t}{(2\pi)^2}
   \frac{(-2s^-)(S^+-s^+)}{(S^--s^-)(S^-+s^-)}
   Tr[\tilde{F}_b(s_t+S_t+q_t)\tilde{F}_c(S_t-s_t-q_t)]\nonumber \\
   \times\left\{\frac{\theta(S^- - s^-)}
   {(S^+ - s^+ - \frac{(S_t -s_t)^{2} -i\epsilon }{2(S^- - s^-)})
   (S^+ + s^+ - \frac{(q_{t}^{2} -i\epsilon )}{2(S^- + s^-)})}\right.
   \nonumber \\
   -\left.\frac{\theta(-S^- + s^-)}
   {(S^+ + s^+ - \frac{(S_t + s_t)^{2} -i\epsilon }{2(S^- + s^-)})
   (S^+ - s^+ - \frac{(q_{t}^{2} -i\epsilon )}{2(S^- - s^-)})}\right\}.
   \label{eq:midstep}
\ee

We will perform the transverse momentum integrations last. For the moment,
let us concentrate on the $s^+$ and $s^-$ integrals. It is more convenient to
do the $s^+$ integral first since this can be done by contour integration. The
$s^+$ dependent integral of the above expression, denoted by $I$, is
\be
   I&\!\!\!=\!\!\!&\int \frac{ds^+}{(2\pi)}\left\{
   \frac{\theta(S^--s^-)}
   {(S^+ - s^+ - \frac{(S_t -s_t)^{2} -i\epsilon }{2(S^- - s^-)})
   (S^+ + s^+ - \frac{(q_{t}^{2} -i\epsilon )}{2(S^- + s^-)})}\right.
   \nonumber \\
   &\!\!\!-\!\!\!&\left.
   \frac{\theta(-S^-+s^-)}
   {(S^+ + s^+ - \frac{(S_t+s_t)^{2} -i\epsilon }{2(S^-+s^-)})
   (S^+-s^+ - \frac{(q_{t}^{2} -i\epsilon )}{2(S^--s^-)})}\right\}
   (S^+ - s^+).
\ee
This integral has a logarithmically divergent piece which can be isolated
by adding and subtracting the term $(S_t -s_t)^{2}/2(S^- - s^-)$ to
the numerator. This divergent piece can be shown to give a constant
independent of the transverse loop momentum and therefore it vanishes upon
dimensional regularization in the transverse direction. The remaining piece
reads as
\be
   I&\!\!\!=\!\!\!& - \int \frac{ds^+}{(2\pi)}
   \left\{
   \frac{\theta(S^--s^-)}
   {(S^+ - s^+ - \frac{(S_t -s_t)^{2} -i\epsilon }{2(S^- - s^-)})
   (S^+ + s^+ - \frac{(q_{t}^{2} -i\epsilon )}{2(S^- + s^-)})}\right.
   \nonumber \\
   &\!\!\!-\!\!\!&\left.
   \frac{\theta(-S^-+s^-)}
   {(S^+ + s^+ - \frac{(S_t+s_t)^{2} -i\epsilon }{2(S^-+s^-)})
   (S^+-s^+ - \frac{(q_{t}^{2} -i\epsilon )}{2(S^--s^-)})}\right\}
   \frac{(S_t-s_t)^2}{2(S^--s^-)}.
   \label{eq:tworplusint}
\ee
Let us investigate the above expression in some detail. The
integrand has two poles in the complex $s^+$ plane. Their
location depends on the signs of $(S^- - s^-)$ and $(S^- + s^-)$. If the two
poles are on the same side of the real axis, then we can close the contour
of integration on the other side of the real axis and the integral vanishes.
So in order to get a non--vanishing result, the two poles must be on opposite
sides of the real axis. Recall that $2S$ is the total or external momentum
flowing into the loop and thus it has to be kept fixed (and finite for a
non--trivially zero loop integral) while working in momentum space. Thus for a
given sign of $S^-$ only one of the two terms in the
integral~(\ref{eq:tworplusint}) above contributes. This can be seen as follows:
the first term in~(\ref{eq:tworplusint}) is
non--zero only if the two conditions
\be
   S^--s^->0,\;\;\;\; S^-+s^->0
   \label{eq:condition1}
\ee
are satisfied, whereas the second term is non--vanishing only if
\be
   S^--s^-<0,\;\;\;\; S^-+s^-<0.
   \label{eq:condition2}
\ee
First, take $S^-$ positive. Then only the first condition~(\ref{eq:condition1})
gives overlapping intervals for $s^-$ namely $S^->s^-$ and $s^->-S^-$ or
$-S^-<s^-<S^-$ whereas the second condition~(\ref{eq:condition2}) does not.
Therefore the second term in equation~(\ref{eq:tworplusint}) can be disregarded
and only the first one is non--vanishing. The opposite is true for $S^-$
negative in which case only the second of the terms in
equation~(\ref{eq:tworplusint}) contributes.

The integral~(\ref{eq:midstep}) however turns out to be independent of the sign
of $S^-$ (as we shall describe below). The reason is the {\it scaling} property
of the integral over $s^-$. This can be understood by recalling that after all,
the overall expression, equation~(\ref{eq:midstep}), is explicitly proportional
to $\delta(2S^-)$ and any term proportional to $S^-$ can be thrown away after
scaling the integration variable $s^-$ by $S^-$. With these remarks in mind,
let us continue working with a definite sign of $S^-$, say, $S^->0$. Performing
the integral~(\ref{eq:tworplusint}) we get
\be
   I=\frac{-i\theta(S^--s^-)\theta(S^-+s^-)(S^-+s^-)(S_t-s_t)^2}
   {S^-[4S^+S^-(1-\xi)(1+\xi) - q_{t}^{2}(1-\xi) - (S_t - s_t)^2(1+\xi)]}
\ee
in terms of which, equation~(\ref{eq:midstep}) is
\be
   \int \frac{d^4s}{(2\pi)^4}
   d^4xd^4ye^{is(x-y)}e^{iS(x+y)}
   \frac{\partial}{\partial y^-}
   D^{ii}_{bc}(x,y)_{(1)}=\nonumber \\
   (-2i)\delta(2S^-)\int \frac{d^2s_t}{(2\pi)^2}\frac{d^2q_t}{(2\pi)^2}
   Tr[\tilde{F}^a(s_t+S_t+q_t)\tilde{F}^b(S_t-s_t-q_t)](S_t -s_t)^2\nonumber \\
   \int_{-S^-}^{S^-} \frac{s^-ds^-}{(S^-\!\! -\! s^-)
   [4S^+(S^-\! +\! s^-)(S^-\!\! -\! s^-)\! -\! q_{t}^{2}(S^-\!\! -\! s^-)
   \!-\!(S_t\! -\! s_t)^2(S^-\! +\! s^-)]}.
\ee
Let us now look at the integration over $s^-$. We scale $s^-$ by $S^-$, that is
we define the variable $\xi = s^-/S^-$.
We notice that after scaling we can make use of the explicit factor
$\delta (2S^-)$ in equation~(\ref{eq:midstep}) and
we can safely throw away any term which is still explicitly proportional
to $S^-$. Thus, the term $[4S^+S^-(1-\xi)(1+\xi)]$ in the denominator of the
above drops. As a result, the overall expression~(\ref{eq:midstep}) becomes
\be
   \int \frac{d^4s}{(2\pi)^4}
   d^4xd^4ye^{is(x-y)}e^{iS(x+y)}
   \frac{\partial}{\partial y^-}
   D^{ii}_{bc}(x,y)_{(1)}=\nonumber \\
   (2i)\delta(2S^-)\int \frac{d^2s_t}{(2\pi)^2}\frac{d^2q_t}{(2\pi)^2}
   Tr[\tilde{F}^a(s_t+S_t+q_t)\tilde{F}^b(S_t-s_t-q_t)](S_t -s_t)^2\nonumber \\
   \int_{-1}^{1} \frac{\xi d\xi}{(1-\xi)
   [q_{t}^{2}(1 - \xi)+(S_t - s_t)^2(1 + \xi)]}.
   \label{eq:overallscale}
\ee

To proceed further, it is convenient to shift  $q_t \rightarrow q_t - s_t$.
Then the arguments of ${F'}^s$ in the trace of the above expression become
independent of $s_t$ and can be taken outside the $s_t$ integral which will
be evaluated next. The $s_t$ integral is a formally divergent integral
and must be regulated. This is done in appendix II using the dimensional
regularization method. We find that
\be
   \int \frac{d^{2\omega} s_t}{(2\pi)^2}
   \frac{(S_t -s_t)^2}{[(q_t - s_t)^2(1-\xi) + (S_t -s_t)^2(1+\xi)]}
   =\frac{\Gamma(-\omega )}{16\pi}\xi(1-\xi)(S_t-q_t)^2,
   \label{eq:divergent}
\ee
which brings expression~(\ref{eq:overallscale}) to read like
\be
   \int \frac{d^4s}{(2\pi)^4}
   d^4xd^4ye^{is(x-y)}e^{iS(x+y)}
   \frac{\partial}{\partial y^-}
   D^{ii}_{bc}(x,y)_{(1)}=
   (2i)\frac{\Gamma(-\omega)}{16\pi}
   \delta(2S^-)\nonumber \\
   \times \int \frac{d^2q_t}{(2\pi)^2}
   Tr[\tilde{F}^a(S_t+q_t)\tilde{F}^b(S_t-q_t)](S_t -q_t)^2
   \int_{-1}^{1} \xi^{2} d\xi .
\ee
The integration over $\xi$ can now be done easily. It just gives a factor of
$2/3$ and the remaining transverse momentum integral can be computed by using
the explicit form of $\tilde{F}$ in terms of the gauge transforms $U$. We also
show this in appendix II where we find that
\be
   \int \frac{d^2q_t}{(2\pi)^2}
   Tr[\tilde{F}_b(S_t+q_t)\tilde{F}_c(S_t-q_t)](S_t -q_t)^2 =
   \frac{g^2}{2}f^{bcd}\tilde{\rho}_d(2S_t)
\ee
with
\be
   \tilde{\rho}_d(2S_t)=\int d^2x_t e^{2iS_tx_t}\rho_d(x_t)
\ee
being the Fourier transform of the charge density with respect to $2S_t$.
Putting everything together, we get finally the result that the expression for
the Fourier transform of the induced current coming from the term proportional
to $\delta_{ij}$ as a function of $2S$ is
\be
   \tilde{j}^+_a(2S)_{(1)}&\!\!\!\equiv \!\!\! & i f_{abc}
   \int \frac{d^4s}{(2\pi)^4}
   d^4xd^4ye^{is(x-y)}e^{iS(x+y)}
   \frac{\partial}{\partial y^-}
   D^{ii}_{bc}(x,y)_{(1)}\nonumber \\
   &\!\!\!=\!\!\!&
   -\frac{\Gamma(-\omega)}{8\pi}g^2
   \tilde{\rho}_d(2S_t)\delta(2S^-) \, .
   \label{eq:contfirstterm}
\ee
Above, we have written only the divergent part of the gamma function when
$\omega \rightarrow 1$ and have used that for $SU(3)$
the product $f_{abc}f_{bcd}\equiv C_A\delta_{ad}$ with $C_A=3$. The remaining
three terms in equation~(\ref{eq:crossterms}) can be evaluated in a similar
fashion. Here we just quote the result
\be
   \tilde{j}^+_a(2S)_{(2)}&=&
   \frac{\Gamma(-\omega)}{8\pi}g^2
   \tilde{\rho}_d(2S_t)\delta(2S^-)\nonumber \\
   \tilde{j}^+_a(2S)_{(3)}&=&
   \frac{\Gamma(-\omega)}{8\pi}g^2
   \tilde{\rho}_d(2S_t)\delta(2S^-)\nonumber \\
   \tilde{j}^+_a(2S)_{(4)}&=&
   4\frac{\Gamma(-\omega)}{8\pi}g^2
   \tilde{\rho}_d(2S_t)\delta(2S^-)\, .
   \label{eq:contrestterms}
\ee
The final result for the Fourier transform of the induced current is obtained
by adding the four terms given by equations~(\ref{eq:contfirstterm})
and~(\ref{eq:contrestterms}) and it becomes
\be
   \tilde{j}^+_a(p) = g^2\Gamma(-\omega)\frac{5}{8\pi}
   \tilde{\rho}_a(p_t)\delta(p^-) \, ,
   \label{eq:totalrenorm}
\ee
where we have renamed $2S \rightarrow p$. Thus the term that modifies the plus
component of the equations of motion in momentum space just becomes
$g\tilde{j}^+_a(p)$.

We proceed to argue that by rewriting this result in terms of an expression
involving the components of the polarization operator, we can absorb the effect
of the loop corrections on the equations of motion, into the renormalization of
the coupling constant. Let us first recall that according to
equations~(\ref{eq:trans}) and~(\ref{eq:alphau}), which are the classical
solutions to the equations of motion, the expression for the zeroth
order background field in momentum space, can be written as
\be
   A^{i(0)}_b(p)&\!\!\!\equiv\!\!\!&\int d^4xe^{ipx}A^{i(0)}_b(x)\nonumber \\
   &\!\!\!=\!\!\!&(2\pi)g\left(\frac{p^i}{p_t^2p^+}\right)
   \tilde{\rho}_b(p_t)\delta(p^-)\, .
\ee
Therefore, notice that $g\tilde{j}^+_a(p)$ can be written as
\be
   g\tilde{j}^+_a(p)=\Pi_{ab}^{+i}(p)A_b^{i(0)}(p)
\ee
with $\Pi_{ab}^{+i}(p)$ given by
\be
   \Pi_{ab}^{+i}(p)= g^2p^+p^i\left(\frac{5\Gamma(-\omega)}{16\pi^2}\right)
   \delta_{ab}
\ee
which is the standard expression for the $\;+i\;$ components of the
polarization
operator in light cone gauge~\cite{leibbrandtreview}. The fact that we recover
this well known result is truly remarkable and is one indicator of the
success of our formalism.

We can now take an {\it ansatz} for the formal solution of the system of
equations~(\ref{eq:set}) to be
\be
   B^-(x)&\!\!\!=\!\!\!&0,\nonumber \\
   B^i(x)&\!\!\!=\!\!\!&\theta(x^-)\alpha^i_R(x_t),\nonumber \\
   \tau\cdot\alpha^i_R(x)&\!\!\!=\!\!\!&\frac{i}{g_R}U(x_t)
   \nabla^iU^{\dagger}(x_t).
   \label{eq:anzatsforB}
\ee
where $\;g_R\;$ is the renormalized coupling constant whose expression is
obtained
through the computation of $\Pi^{\mu\nu}$ and will be given explicitly by
\be
   g_{R}^{2}=Z_3g^2,\;\;\;\;\;
   Z_3=\left(1+\frac{11g^2}{16\pi^2(1-\omega)}\right)\, .
\ee
This in turn means, according to~(\ref{eq:anzatsforB}), that the field $B$ gets
renormalized by the inverse of the constant that renormalizes $g$:
\be
   B^{i}(x)=Z^{-1/2}A^{i(0)}(x).
\ee

The above exercise has taught us the important lesson that the modifications to
the background field introduced by the quantum fluctuations do not induce extra
terms in the expression for the distribution function~(\ref{eq:MAINRES}).
Further, their effect can be included by replacing the coupling constant $g$ by
the renormalized coupling constant $g_R$.

\section{Summary}

We have presented in this paper an expression for the quantum corrections to
the Weizs\"acker--Williams gluon distribution at small x valid to all orders in
the parameter $\alpha_s\mu$. We used this expression to compute explicitly the
leading $\ln(1/x)$ and $\ln(k_t)$ terms in the momentum regime
$\alpha_{s}\mu\ll k_{t}$.  We have shown that the perturbative approach
introduces a series expansion parameter $\alpha_s \ln(1/x)$ which is large and
thus, forces us to devise a method to sum up the leading contributions to all
orders in that expansion parameter. Nevertheless, the present result already
signals that at small x values the gluon distribution function will be modified
significantly from the $1/(xk_t^2)$ behavior.

We wish to emphasize that our central result, equation~(\ref{eq:MAINRES}),
contains in principle the information about the quantum correction to the
classical distribution to all orders in the parameter $\alpha_s\mu$.

We have found that the only effect the quantum corrections have on the
classical background field can be absorbed into the renormalization of the
field and the running of the coupling constant.

We have not addressed the issue of summing up the perturbative series in this
paper. In the weak coupling limit, this is equivalent to solving an integral
equation for virtual corrections to the gluon propagator. Another issue we
would like to address is whether  we can relax some of the constraints in the
model. In particular, we need to pay attention to the restriction set by the
necessity of having a large (perhaps too large) $A$ nucleus in order to compare
our predictions to experimental data. These issues will be addressed in a
future work.

\section*{Appendix I:  The propagator in momentum space}

In this appendix, we shall Fourier transform the small fluctuations propagator
to momentum space.  This result will be useful for computing
the distribution function and may also be useful for other computations.
We will first consider the propagator without averaging over all
possible color orientations of the source fields.

We shall derive two representations of the propagator. One representation
will include an integration over the variable $p^+$. This representation
will be useful for computing distribution functions.  We will also present
a second representation where we have performed this integration over $p^+$.
Finally, we will present a result for the propagator after averaging over
the colors of the external field.

We recall that in coordinate space the propagator has the form
\be
G_{ij}^{\alpha\beta ;\alpha^\prime \beta^\prime}(x,y)
\!\!\!\!\!& = &\!\!\!\! - \! \int \!\!{{d^4p}\over {(2\pi)^4}}
{{e^{ip(x-y)}}\over {p^2-i\epsilon}}
\Bigg\{\! \bigg[\delta_{ij}\!+\!{p_ip_j\over {p^-p^+}}(2e^{ip^+(x^-\! - y^-)}
\!-\! e^{-ip^+y^-}\!-\! e^{ip^+x^-})\bigg]\nonumber \\
&\times &   \bigg[\theta(-x^{-})\theta(-y^{-})\tau_{a}^{\alpha\beta}
\tau_{a}^{\alpha^\prime \beta^\prime} + \theta(x^-)\theta(y^-)\,
F_{a}^{\dagger\alpha\beta}(x_{t})
F_{a}^{\dagger\alpha^\prime \beta^\prime}(y_{t})
\bigg]\nonumber \\
\!\!\!\!\!&+ &\!\!\!\!
 \theta(-x^-) \theta(y^-)
\int \!{{d^2 q_t} \over {(2\pi)^2}} d^2 z_t \,\, e^{i(q^{+}-p^{+})y^{-}}
e^{i(p_t-q_t)(y_t-z_t)} \nonumber \\
\!\!\!\!\!&\times &\!\!\!\!
F_{a}^{\dagger\alpha \beta}(z_{t})
F_{a}^{\dagger\alpha^\prime  \beta^\prime}\!(y_{t})\bigg[\delta_{ij} +
\frac{p_{i}p_{j}}{p^-p^{+}}
(e^{ip^{+}x^{-}}\!-1)\nonumber \\
\!\!\!\!\!& + &\!\!\!\!\!\!   \frac{q_{i}q_{j}}{p^-q^{+}}
(e^{-iq^{+}y^{-}}\!-1)
+ \frac{p_{i}q_{j}p_{t}\cdot q_{t}}
{(p^{-}p^{+})(p^{-}q^{+})}(e^{ip^{+}x^{-}}\!-1)
(e^{-iq^{+}y^{-}}\!-1)\bigg]\nonumber \\
\!\!\! & + &\!\!\!\!\!
\theta(x^-) \theta(-y^-)
\int\! {{d^2 q_t} \over {(2\pi)^2}} d^2 z_t \,\,
 e^{i(p_t-q_t)(z_t - x_t)} e^{-i(q^{+}-p^{+})x^{-}}\nonumber \\
\!\!\!\!\!&\times &\!\!\!\! F_{a}^{\dagger\alpha \beta}(x_{t})
F_{a}^{\dagger\alpha^\prime  \beta^\prime}\!(z_{t})\bigg[\delta_{ij} +
\frac{p_{i}p_{j}}{p^{-}p^{+}}(e^{-ip^{+}y^{-}}\!\!-1)
 +  \frac{q_{i}q_{j}}{p^{-}q^{+}}(e^{iq^{+}x^{-}}\!-1)\nonumber \\
\!\!\!\!\!& + &\!\!\!\!\!\!
\frac{q_{i}p_{j}p_{t}\cdot q_{t}}{(p^{-}p^{+})(p^{-}q^{+})}
(e^{-ip^{+}y^{-}}\!-1)
(e^{iq^{+}x^{-}}\!-1)\bigg]\Bigg\} \, ,
\label{eq:GF1}
\ee
where $q^{+}=p^{+}+ \frac{q_{t}^{2}-p_{t}^{2}}{2p^{-}}$ and
\be
   F_{a}^{\dagger\alpha \beta}(x_{t}) =\bigg( U(x_{t})\tau_{a}U^{\dagger}
   (x_{t})\bigg)^{\alpha\beta}.
\ee

We now wish to define the Fourier transformed Green's function
\be
	G^{\alpha \beta; \alpha^\prime \beta^\prime}_{ij} (k,k^\prime)
 = \int d^4xd^4y e^{-ikx+ik^\prime y}
G^{\alpha \beta; \alpha^\prime \beta^\prime}_{ij} (x,y) \, ,
\ee
which we can divide into four pieces as
\be
G^{\alpha \beta; \alpha^\prime \beta^\prime}_{ij} (k,k^\prime) & = &
G^{\alpha \beta; \alpha^\prime \beta^\prime}_{--ij} (k,k^\prime) +
G^{\alpha \beta; \alpha^\prime \beta^\prime}_{++ij} (k,k^\prime)\nonumber \\
& + &
G^{\alpha \beta; \alpha^\prime \beta^\prime}_{+-ij} (k,k^\prime) +
G^{\alpha \beta; \alpha^\prime \beta^\prime}_{-+ij} (k,k^\prime)
\ee
In this equation, the first index $\pm$ refers to the index of
$\theta (\pm x^-)$ and the second to $\theta (\pm y^-)$ in the definition
of the coordinate space Green's function.

Let us first consider the $--$ component
\be
G^{\alpha \beta; \alpha^\prime \beta^\prime}_{--ij} (k,k^\prime)
&\!\!\! = \!\!\! &
- \int {{d^4p} \over {(2\pi)^4}} d^4xd^4y e^{i(p-k)x-i(p-k^\prime)y}
\theta(-x^-) \theta (-y^-) \tau_a^{\alpha \beta} \tau_a^{\alpha^\prime
\beta^\prime} {1 \over {p^2-i\epsilon}} \nonumber \\
&\!\!\!\times\!\!\!& \left\{ \delta_{ij}
+{{p_ip_j} \over {p^-p^+}} (2e^{ip^+(x^--y^-)} - e^{-ip^+y^-}
-e^{ip^+x^-}) \right\} \, .
\ee
We can perform the integrations over $p^-$, $p_t$, $\vec{x}$
and $\vec{y}$ to obtain,
\be
G^{\alpha \beta; \alpha^\prime \beta^\prime}_{--ij} (k,k^\prime)
= - (2\pi) \delta(k^- - k^{\prime -}) (2\pi)^2 \delta^{(2)}(k_t - k^\prime_t)
\tau_a^{\alpha \beta} \tau_a^{\alpha^\prime \beta^\prime}
\Delta_{--}(k,k^\prime)
\ee
where
\be
\Delta_{--}(k,k^\prime) & = & \int dx^-dy^- {{dp^+} \over {2\pi}}
\frac{\theta (-x^-) \theta (-y^-) e^{-i(p^+-k^+)x^- + i (p^+-k^{\prime +})y^-}}
{k_t^2 - 2p^+k^- - i \epsilon}\nonumber \\
&\times&\Bigg\{ \delta_{ij}
+{{k_ik_j} \over {k^-p^+}} (2e^{ip^+(x^--y^-)} - e^{-ip^+y^-}
-e^{ip^+x^-}) \Bigg\}
\ee

Now let us do the $x^-$ and $y^-$ integrations.  In performing these
integrations, $i\epsilon$ factors will appear which will guarantee
the convergence of the integrals at infinity.  We find
\be
	 \Delta_{--}(k,k^\prime) & = & \int {{dp^+} \over {2\pi}}
\,{1 \over {k_t^2 - 2 p^+k^- -i\epsilon}}\, {1 \over {p^+-k^++i\epsilon}}
\,{1 \over {p^+-k^{\prime+}-i\epsilon}} \nonumber \\
&\times& \left\{ \delta_{ij} + {{k_ik_j} \over {k^-(k^+-i\epsilon)
(k^{\prime +} + i\epsilon)}} (2p^+ - k^+ -k^{\prime +}) \right\}
\ee
Finally, we can perform the integration over $p^+$ to find
\be
	\Delta_{--}(k,k^\prime)& =& {i \over {k^{\prime +} - k^+ +i\epsilon}}
\,{1\over {k^{\prime 2} -i\epsilon}}\,\Bigg\{ \theta (k^-) \Big( \delta_{ij}
 \nonumber \\
&+& {{k_ik_j} \over {k^-(k^+-i\epsilon)
(k^{\prime +} + i\epsilon)}} (k^{\prime +} -k^+) \Big) +
k,k^\prime \rightarrow -k^\prime,-k \Bigg\}
\ee

Now, to fully define this Green's function, we must specify the nature
of the singularity at $k^- = 0$.  In the last equation, we would like
to use the Leibbrandt--Mandelstam prescription on $1/(k^- +i\epsilon /k^+)$
whenever we have the combination $1/k^-k^+$ and
$1/(k^-+i\epsilon /k^{\prime +})$ whenever we have the combination
$1/k^-k^{\prime +}$.  We would like to go further however
and define $1/k^+k^-$ as $1/(k^-k^+ + i\epsilon)$.  This can be done
as follows: we use $1/(k^+-i\epsilon) = 2\pi i \delta (k^+) +
1/(k^++i\epsilon)$
whenever we have the constraint that $k^- > 0$, and a similar modification
when $k^- < 0$ for ${k^\prime}^+$.

On the other hand, it is more difficult to implement the Leibbrandt-Mandelstam
prescription in the expression which involves the integral over $p^+$.
However, we will only use this result when both $k^+$ and $k^{\prime +}$ have
the same sign and are non-zero. In this case the Leibbrandt-Mandelstam
prescription is unambiguous.

Our results for the two representations are therefore
\be
	G^{\alpha \beta; \alpha^\prime \beta^\prime}_{--ij} (k,k^\prime)
 & = & -2\pi i \delta(k^- - k^{-\prime}) (2\pi)^2 \delta^{(2)}
(k_t - k_t^\prime) \tau_a^{\alpha \beta} \tau_a^{\alpha^\prime \beta^\prime}
{1\over {k^{\prime 2} -i\epsilon}}
\nonumber \\
 &\times&\Bigg\{ {1 \over {k^{\prime +} - k^+ +i\epsilon}}
\Big[ \theta (k^-) \left( \delta_{ij} + {{k_ik_j} \over {k^-k^+
k^{\prime+}}} (k^{\prime +} -k^+) \right) \nonumber \\
&+& k,k^\prime \rightarrow -k^\prime,-k \Big]
+2\pi i \delta (k^+) \theta (k^-) {{k_ik_j} \over {
k^-k^{\prime+}}}\nonumber\\
&+& k,k^\prime \rightarrow -k^\prime,-k \Bigg\}
\ee
Here the $k^+$, $k^{\prime +}$ and $k^-$ singularities are treated using
the Leibbrandt-Mandelstam prescription.

For $k^+$ and $k^{\prime +}$ both non-zero and of the same sign, we have the
representation
\be
	G^{\alpha \beta; \alpha^\prime \beta^\prime}_{--ij} (k,k^\prime)
 & = & -2\pi \delta(k^- - k^{-\prime}) (2\pi)^2 \delta^{(2)}
(k_t - k_t^\prime) \tau_a^{\alpha \beta} \tau_a^{\alpha^\prime \beta^\prime}
\nonumber \\
&\times& \int {{dp^+} \over {2\pi}}
{1 \over {k_t^2 - 2 p^+k^- -i\epsilon}}\, {1 \over {p^+-k^++i\epsilon}}\,
{1 \over {p^+-k^{\prime+}-i\epsilon}} \nonumber \\
&\times& \left\{ \delta_{ij} + {{k_ik_j} \over {k^-k^+
k^{\prime +}}} (2p^+ - k^+ -k^{\prime +}) \right\}\, ,
\ee
where the $k^-$ singularity is treated using the Leibbrandt-Mandelstam
prescription.

The evaluation of the remaining contributions to the Green's functions can be
done by the same methods as above. There is nothing really new in the analysis
except that it is longer and more involved.   The subtlety is in the treatment
of the singularities in the $k^\pm$ and $k^{\prime +}$ variables. This has been
discussed above and is treated as such. The results are
\be
G^{\alpha \beta; \alpha^\prime \beta^\prime}_{++ij} (k,k^\prime)
&\!\!\! = \!\!\!& -2\pi i \delta (k^- - k^{-\prime})
\int {{d^2p_t} \over {(2\pi)^2}}
F_a^{\dagger \alpha \beta}(p_t-k_t)
F_a^{\dagger \alpha^\prime \beta^\prime}(k_t^\prime -p_t) \nonumber \\
&\!\!\!\times\!\!\!& \Bigg\{ {1 \over {k^+ - k^{\prime +} +i\epsilon}}
\Bigg[ \left( \delta_{ij} +
{{p_ip_j} \over {k^-k^+k^{\prime +}}}(k^+-k^{\prime +}) \right)
{ \theta (k^-)\over {p_t^2 - 2k^-k^+ - i\epsilon}} \nonumber \\
&\!\!\!+\!\!\!&   k, k^\prime \rightarrow -k^\prime k \Bigg]
+2\pi i \delta (k^{\prime +}) \theta (k^-) {{p_ip_j} \over {k^-k^+}}
{1 \over {p_t^2-2k^-k^+-i\epsilon}} \nonumber \\
&\!\!\!+\!\!\!& k,k^\prime \rightarrow -k^\prime, -k
\Bigg\}\, .
\ee
For $k^+$ and $k^{\prime +}$ both of the same sign and non-zero, the above is
equivalent to
\be
G^{\alpha \beta; \alpha^\prime \beta^\prime}_{++ij} (k,k^\prime)
& = & -2\pi  \delta (k^- - k^{-\prime}) \int {{dp^+d^2p_t} \over {(2\pi)^3}}
F_a^{\dagger \alpha \beta}(p_t-k_t)
F_a^{\dagger \alpha^\prime \beta^\prime}(k_t^\prime - p_t)
\nonumber \\
&\times&{ 1 \over {p_t^2-2p^+k^--i\epsilon} }\,
{1 \over {p^+-k^+-i\epsilon}}\,{1 \over {p^+-k^{\prime +} + i\epsilon}}
\nonumber \\
&\times& \left( \delta_{ij} + {{p_ip_j} \over {k^-k^+k^{\prime +}}}
(2p^+-k^+-k^{\prime +}) \right) \, .
\ee

We finally also obtain an expression for $G_{+-}$.  It turns out that in this
expression, no restrictions on the values of $k^+$ and $k^{\prime +}$ are
needed to get the singularities in $1/k^\pm$ or $1/k^{\prime +}$ into the
Leibbrandt--Mandelstam form. The results are
\be
G^{\alpha \beta; \alpha^\prime \beta^\prime}_{+-ij} (k,k^\prime)
& = & 2\pi i \delta (k^- - k^{-\prime}) 2k^- \theta (k^-)
{1 \over {k^{\prime 2}-i\epsilon}} \int {{d^2p_t} \over {(2\pi)^2}}
F_a^{\dagger \alpha \beta}(p_t-k_t)
\nonumber \\
&\times&
F_a^{\dagger \alpha^\prime \beta^\prime}(k_t^\prime -p_t)
{1 \over {p_t^2-2k^-k^+ -i\epsilon}} \Bigg\{ \delta_{ij}
  - {{k^\prime_ik^\prime_j} \over {k^-k^{\prime +}}}
-{{p_ip_j} \over {k^-k^+}} \nonumber \\
&+& {{p_ik^\prime_j p_t \cdot k^\prime} \over {(k^-k^+)(k^-k^{\prime +})}}
\Bigg\}  \, .
\ee
We also have the equivalent integral representation where
\be
	G^{\alpha \beta; \alpha^\prime \beta^\prime}_{+-ij} (k,k^\prime)
& = & 2\pi  \delta (k^- - k^{-\prime}) \int {{dp^+d^2p_t} \over {(2\pi)^3}}
F_a^{\dagger \alpha \beta}(p_t-k_t)
F_a^{\dagger \alpha^\prime \beta^\prime}(k_t^\prime -p_t) \nonumber \\
&\times&  {1 \over {k_t^2 -2k^-p^+ -i\epsilon}}\, {1 \over {q^+-k^+-i\epsilon}}
\,{1 \over {p^+-k^{\prime +} -i\epsilon}} \nonumber \\
&\times& \Bigg\{ \delta_{ij} -{{k^\prime_i k^\prime_j} \over
{k^-k^{\prime +}}}
-{{p_ip_j} \over {k^-k^+}} + {{p_ik^\prime_j p_t \cdot k^\prime_t}
\over {(k^-k^+) (k^-k^{\prime +})}}\Bigg\} \, .
\ee
In this equation,
$q^+ = p^+ + {{p_t^2-k^{\prime 2}} \over {2k^-}}$.

The expression for $G_{-+}$ is
\be
	G^{\alpha \beta; \alpha^\prime \beta^\prime}_{+-ij} (k,k^\prime)
= 	G^{\alpha \beta; \alpha^\prime \beta^\prime}_{-+ij} (-k^\prime,-k)
\ee

We now want to convert these expressions for the propagator from the
matrix basis to the component basis.  To do this, we make the
transformation
\be
	(U(x)\tau^c U^\dagger (x)) (U(y) \tau^c U^\dagger (y))
\rightarrow 4 (Tr\, \tau^a U(x) \tau^c U^\dagger (x)) (Tr\,
\tau^b U(y) \tau^c U^\dagger (y) )
\ee
We then use the identity
\be
	\tau^c_{\alpha \beta} \tau^c_{\alpha^\prime \beta^\prime}
= {1 \over 2} \left( \delta_{\alpha \beta^\prime} \delta_{\alpha^\prime \beta}
 - {1 \over N_c} \delta_{\alpha \beta} \delta_{\alpha^\prime \beta^\prime}
\right)\, ,
\ee
which results in following transformation for the definition of the propagator
\be
	F^{\dagger a}_{\alpha \beta} (x)
F^{\dagger a}_{\alpha^\prime  \beta^\prime} (y)
\rightarrow 2 \,Tr \, F^a(x) F^b(y)
\ee

We now want to proceed to derive formulae for the propagator
which has been averaged over all values of the color charges of
the valence quarks.  To do this we define
\be
< Tr \,U^\dagger (x) \tau^a U(x) U^\dagger (y) \tau^b U(y) >
= {1 \over 2} \Gamma^{ab}(x-y) \, .
\ee
Notice that
\be
	\Gamma^{ab} (0) = \delta^{ab}
\ee

We now define
\be
	< G^{ab}_{ij} (k,k^\prime ) > = (2\pi ) \delta (k^- - k^{\prime -} )
(2\pi )^2 \delta (k_t - k^\prime_t) D_{ij}^{ab} (k,k^\prime) \, .
\ee
As before, we can write $D_{ij}^{ab}$ as
\be
	D_{ij}^{ab} = D_{-- ij}^{ab}(k,k^\prime) + D_{++ ij}^{ab}(k,k^\prime)
+  D_{+- ij}^{ab}(k,k^\prime) +  D_{-+ ij}^{ab}(k,k^\prime)\, .
\ee

Using the above substitutions we find that
\be
	D_{--ij}^{ab} & = & -\delta^{ab} \int {{dp^+} \over {2\pi}}
\, {1 \over {k_t^2 - 2 p^+ k^- -i\epsilon}}\,
{1 \over {p^+-k^+ +i\epsilon}}\, { 1 \over {p^+-k^{\prime +} -i\epsilon}}\,
\nonumber \\
&\times&\left[ \delta_{ij} + {{k_ik_j} \over {k^-k^+k^{\prime +}}}
(2p^+-k^+-k^{\prime +}) \right]\, ,
\ee
which is valid for $k^+,k^{\prime +}$ nonzero and $sign(k^+) = sign(k^{\prime
+} )$ . The expression with the $p^+$ integral completed is
\be
	D_{--ij}^{ab} & = & -i \delta^{ab} \Bigg\{ \theta(k^-)
{1 \over k^{\prime 2}}
 \Bigg[ {1 \over {k^{\prime +} -k^+ +i\epsilon }}
\left( \delta_{ij} + {{k_ik_j} \over {k^-k^+k^{\prime +}}}
(k^{\prime +} - k^+) \right)
\nonumber \\
&+&2\pi i \delta (k^+) {{k_ik_j} \over {k^-k^{\prime +}}} \Bigg]
+ k,k^\prime \rightarrow -k^\prime, -k \Bigg\} \, .
\ee

For $D_{++}$ we obtain
\be
	D_{++ij}^{ab} & = & -\int {{dp^+d^2p_t} \over {(2\pi )^3}}
\Gamma^{ab}(p_t-k_t)
\, {1 \over {p_t^2 - 2p^+k^- -i\epsilon}} \nonumber \\
&\times&{1 \over {p^+-k^+-i\epsilon}}\,{1 \over {p^+-k^{\prime +} + i\epsilon
}}  \left[ \delta_{ij} + {{p_ip_j} \over {k^-k^+k^{\prime +}}}
(2p^+-k^+-k^{\prime +}) \right] \, ,
\ee
which is valid for $k^+,k^{\prime +}$ nonzero and $sign(k^+) = sign(k^{\prime
+} )$ . After completing the $p^+$ integral,
\be
	D_{++ij}^{ab} & = & \int {{d^2p_t} \over {(2\pi )^2}}
\Gamma^{ab}(p_t-k_t)
\Bigg\{ \theta (k^-) {1 \over {p_t^2 - 2k^-k^+ -i\epsilon}}
\nonumber \\ &\times&\Bigg[{1 \over {k^+-k^{\prime +}}} \left( \delta_{ij} +
{{p_ip_j} \over {k^-k^+k^{\prime +}}} (k^+-k^{\prime +} ) \right) + 2\pi i
\delta (k^{\prime +}) {{p_ip_j} \over {k^-k^+}} \Bigg]\nonumber \\
&+& k,k^\prime \rightarrow -k^\prime, -k \Bigg\} \, .
\ee
We also have
\be
	D_{+-ij}^{ab} & = & \int {{dp^+d^2p_t} \over {(2\pi)^3}}
\Gamma^{ab}(p_t-k_t)
\,{1 \over {k_t^2-2k^-p^+-i\epsilon}}\, {1 \over {q^+-k^+-i\epsilon}}\,
{1 \over {p^+-k^{\prime +} +i\epsilon}}\,
\nonumber \\
&\times&\Bigg[\delta_{ij} - {{k_ik_j} \over {k^-k^{\prime +}}}
-{{p_ip_j} \over {k^-k^+}} + {{p_i k_j p_t \cdot k_t} \over {(k^-k^+)
(k^-k^{\prime +})}} \Bigg] \, ,
\ee
where, again,
$q^+ =  p^+ + {{p_t^2-k_t^2} \over {2k^-}}$.
Doing the integral over $p^+$ gives
\be
	D_{+-ij}^{ab} & = & i 2k^- \theta (k^-) {1 \over {k^{\prime 2}}}
\int {{d^2p_t} \over {(2\pi )^2}}\, \Gamma^{ab} (p_t-k_t)
{1 \over {p_t^2 - 2k^-k^+ -i\epsilon}}
\nonumber \\ &\times&
\Bigg[\delta_{ij} - {{k_ik_j} \over {k^-k^{\prime +}}}
-{{p_ip_j} \over {k^-k^+}} + {{p_i k_j p_t \cdot k_t} \over {(k^-k^+)
(k^-k^{\prime +})}} \Bigg] \, .
\ee

Finally, the expression for $D_{-+}$ is given by
\be
	D_{-+ij}^{ab} (k,k^\prime) = D^{ba}_{+-ji} (-k^\prime,-k) \, .
\ee

\section*{Appendix II}

In this appendix, we will explicitly evaluate two of the integrals we
use in the computation of the one loop corrections to the classical background
field. The first integral we will evaluate is
\be
 I_1= \int \frac{d^2s_t}{(2\pi)^2}
   \frac{(S_t -s_t)^2}{[(q_t - s_t)^2(1-\xi) + (S_t -s_t)^2(1+\xi)]}.
\ee
Let us first shift $s_t \rightarrow s_t+S_t$ and write the above as
\be
I_1= \int \frac{d^2s_t}{(2\pi)^2}
   \frac{s_{t}^{2}}{[(s_t+S_t-q_t)^2(1-\xi) + s_{t}^{2}(1+\xi)]}\, .
   \label{eq:shift}
\ee
The denominator in the integral can be written as
\be
   2s_{t}^{2} + (2s_t\cdot(S_t-q_t) + (S_t-q_t)^2)(1-\xi).
   \label{eq:denominator}
\ee
Performing the change of variables
\be
   v_t&=&s_t + \frac{(1-\xi)}{2}(S_t-q_t)\nonumber \\
   d^2v_t&=&d^2s_t\nonumber \\
   v_t^2&=&s_{t}^{2} + s_t\cdot(S_t-q_t)(1-\xi)
         + \frac{(1-\xi)^2}{4}(S_t-q_t)^2
   \label{eq:changevar}
\ee
Eq.~(\ref{eq:denominator}) can be re--written as
\be
   2[v_t^2 + \frac{(1-\xi^2)}{4}(S_t-q_t)^2]
   \label{eq:newdenominator}
\ee
and the integral~(\ref{eq:shift})
\be
I_1= \int \frac{d^2v_t}{(2\pi)^2}
   \frac{[v_t - \frac{(1-\xi)}{2}(S_t-q_t)]^2}
        {2[v_t^2 + \frac{(1-\xi^2)}{4}(S_t-q_t)^2]}.
   \label{eq:newshift}
\ee
Written in this form, we can drop the linear term in $v_t$ and the contributing
terms to~(\ref{eq:newshift}) will be
\be
I_1   &\rightarrow&\int \frac{d^2v_t}{(2\pi)^2}
        \frac{v_t^2}{2[v_t^2 + \frac{(1-\xi^2)}{4}(S_t-q_t)^2]}\nonumber \\
   & +&\int \frac{d^2v_t}{(2\pi)^2}
        \frac{(1-\xi)^2/4(S_t-q_t)^2}{2[v_t^2 +
\frac{(1-\xi^2)}{4}(S_t-q_t)^2]}.
   \label{eq:twoterms}
\ee
The first and second terms in expression~(\ref{eq:twoterms}) are
quadratically and logarithmically divergent. To regulate the divergences we
compute them by dimensional regularization. We thus write~(\ref{eq:twoterms})
as
\be
I_1&=&   \frac{1}{2(2\pi)^2} \Bigg( \frac{2\pi^2}{\Gamma(d/2)} \Bigg)
   \Bigg\{
   \int_{0}^{\infty}
   dv_t \frac{v_t^{d+1}}{[v_t^2 + \frac{(1-\xi^2)}{4}(S_t-q_t)^2]}\nonumber \\
&+&
   \int_{0}^{\infty}
    dv_t \frac{v_t^{d-1}(1-\xi)^2/4(S_t-q_t)^2}
            {[v_t^2 + \frac{(1-\xi^2)}{4}(S_t-q_t)^2]}
   \Bigg\}.
   \label{eq:dimreg}
\ee
To evaluate the integrals in the above expression we use the well known
formula
\be
   \int_{0}^{\infty}
   \frac{u^{\beta}du}{(u^2+C^2)^\alpha} =
   \frac{\Gamma(\frac{1}{2}(1+\beta))\Gamma(\alpha-\frac{1}{2}(1+\beta))}
        {2(C^2)^{\alpha-(1+\beta)/2}\Gamma(\alpha)}
   \label{eq:identity}
\ee
by means of which~(\ref{eq:dimreg}) becomes
\be
I_1= (\frac{d}{4})(\frac{1}{2\pi})^2 2\pi^{d/2}\Gamma(-d/2)
   (\frac{x}{1+x})[(1-\xi^2)(S_t-q_t)^2/4]^{d/2}.
   \label{eq:almostresult}
\ee
We are interested in the divergent part of this expression when
$d\rightarrow 2$ thus for that we take $d=2$ everywhere except in the argument
of $\Gamma$. We obtain finally
\be
I_1= \frac{\Gamma(-\frac{d}{2})}{16\pi}2\xi(1-\xi)(S_t-q_t)^2,
\ee

Next, we want to compute the following integral
\be
I_2=\int \frac{d^2q_t}{(2\pi)^2}
   Tr[F^a(S_t+q_t)F^b(S_t-q_t)][(S_t -q_t)^2 - (S_t +q_t)^2].
   \label{eq:chargeprop}
\ee
For this purpose let us write $\tilde{F}$ in its explicit form in terms of $U$.
\be
   \tilde{F}^a(p_t)=\int d^2x_t e^{ip_tx_t} U^{\dagger}(x_t)\tau^aU(x_t),
   \label{eq:tildeF}
\ee
and thus the integrand in~({\ref{eq:chargeprop}) can be written as
\be
   \int d^2x_td^2y_t e^{i(S_t+q_t)x_t}e^{i(S_t-q_t)y_t}
      [(S_t-q_t)^2 - (S_t+q_t)^2]\nonumber \\
   Tr[U^{\dagger}(x_t)\tau^aU(x_t)U^{\dagger}(y_t)\tau^bU(y_t)].
   \label{eq:FTint}
\ee
The factors $(S_t\pm q_t)^2$ are to be interpreted as derivatives acting on the
corresponding exponential. Integrating by parts, ignoring the surface terms and
with the help of Eq.~(\ref{eq:FTint}) we can write Eq.~(\ref{eq:chargeprop}) as
\be
I_2&=& -\int\frac{d^2q_t}{(2\pi)^2}d^2x_td^2y_t
e^{i(S_t+q_t)x_t}e^{i(S_t-q_t)y_t}    \nonumber \\
   &\times&\{Tr[U^{\dagger}(x_t)\tau^aU(x_t)
           \nabla^2(U^{\dagger}(y_t)\tau^bU(y_t))]\nonumber \\
   &-&Tr[\nabla^2(U^{\dagger}(x_t)\tau^aU(x_t))
           U^{\dagger}(y_t)\tau^bU(y_t)]\}.
\ee
Performing the $q_t$
and $y_t$ integrals, Eq.~(\ref{eq:chargeprop}) becomes
\be
I_2&=& -\int d^2x_t e^{2iS_tx_t}
   \Bigg\{ Tr[U^{\dagger}(x_t)\tau^aU(x_t)
           \nabla^2(U^{\dagger}(x_t)\tau^bU(x_t))]\nonumber \\
     &-& Tr[\nabla^2(U^{\dagger}(x_t)\tau^aU(x_t))
           U^{\dagger}(x_t)\tau^bU(x_t)]
   \Bigg\}\, .
\label{eq:about}
\ee
Since the gauge transformations $U$ and the charge density are related by
Eq.~(\ref{eq:ufield}), one can prove the following identity
\be
   Tr[U^{\dagger}\tau^aU
           \nabla^2(U^{\dagger}\tau^bU) -
   \nabla^2(U^{\dagger}\tau^aU)
           U^{\dagger}\tau^bU]= -\frac{g^2}{2}\Bigg(f^{abc}\rho_c(x_t) -
f^{bac}\rho_c(x_t)\Bigg)\!.
\label{eq:tracediff}
\ee
Using the antisymmety of $f^{abc}$ and plugging the above back into
Eq.~(\ref{eq:about}) we finally obtain the result
\be
I_2= g^2f^{abc}\int d^2x_t e^{2iS_tx_t} \rho_c(x_t) \, ,
\ee
which is the Fourier transform of the charge density with respect to $2S_t$.

\section*{Acknowledgments}

Two of the authors, A.A and J.J.M., would like to thank R. Rodriguez and
R. Madden for useful discussions.
Research supported by the U.S. Department of Energy under grants No.
DOE High Energy DE--AC02--83ER40105, No. DOE Nuclear DE--FG02--87ER--40328,
No. DOE Nuclear DE--FG06--90ER--40561, and by the DGAPA/UNAM/M\'exico.

\newpage

\section*{Figure captions}

Figure 1: a) Coupling of the classical background field to the external
source. The wavy line represents the background field which is of
order~$O\,(\frac{1}{g})$. The external source is shown by a cross and to the
lowest order in the weak coupling regime it is $O\,(\frac{\mu g^2}{k_t})$.
b) Correlation of two classical background fields where the broken wavy
line means that the momentum~$k_t$ is not integrated over.
\\
Figure 2: a) The perturbative expansion of the gluon propagator in the
presence of the classical background field in terms of the coupling constant
$g$. b) The gluon propagator expanded to the second order in g.
\\
Figure 3: The correction to the Weizs\"{a}cker--Williams distribution
function to the lowest order in the weak coupling regime.
\\
Figure 4: Modification of the classical background field due to
quantum fluctuations.

\end{document}